# Two-stage Information Spreading Evolution on The Control Role of Announcements

Jinhu Ren[1], Fuzhong Nian*[1], and Xiaochen Yang[1]

*Abstract*—Modern social media networks have become an important platform for information competition among countries, regions, companies and other parties. This paper utilizes the research method of spread dynamics to investigate the influence of the control role of announcements in social networks on the spreading process. This paper distinguishes two spreading phases using the authentication intervention as a boundary: the unconfirmed spreading phase and the confirmed spreading phase. Based on the actual rules of spreading in online social networks, two kinds of verification results are defined: true information and false information. The Two-stage information spreading dynamics model is developed to analyze the changes in spreading effects due to different validation results. The impact of the intervention time on the overall spread process is analyzed by combining important control factors such as response cost and time-sensitivity. The validity of the model is verified by comparing the model simulation results with real cases and the adaptive capacity experiments. This work is analyzed and visualized from multiple perspectives, providing more quantitative results. The research content will provide a scientific basis for the intervention behavior of information management control by relevant departments or authorities.

*Index Terms*—Complex systems, opinion dynamics, social network dynamics, network optimization and control.

## I. Introduction

### A. Background and Research Status

Social media software has become the main medium of information dissemination in the modern social system [1, 2]. It has also become an important platform for public information management. Research related to complex networks [3] has led to breakthroughs in numerous practical areas such as information spreading and traffic prediction [4-7]. The study of spreading dynamics on networks is one of the important directions in the field of complex networks research [8-11]. Information spreading models have been of great relevance in exploring message diffusion [12, 13], opinion control [14, 15], etc. The emergence of new social media [16] networks due to the development of modern information technology have further changed people's original lifestyles [17] and work activities [18]. People get a wide variety of news and information through social software and become communicators and discussants in social networks [19-22]. In the modern world, it is clear that social media networks have become an important platform for information competition among countries, regions, companies, and other parties [23-26]. How to ensure the widespread dissemination of true information and control the scale of spreading false information has become a hot topic of current research [27-31]. This paper further explores the information control role generated by the government or authority in the field of public information management using relevant research in spreading dynamics. Through modeling and analysis, the optimal intervention time as well as the effectiveness of information control are explored. We provide a scientific basis for optimal decision time for the government or authority. In the following, the way information is controlled is uniformly defined as an announcement.

Current research generally agrees that building information transmission models [32] based on classical infectious disease models is a feasible approach. Kumar et al. [33] established the SEI model (Susceptible, Exposed-Infected, SEI) model based on the SI (Susceptible-Infective, SI) model to analyze the spreading pattern of individual messages in social networks. Li et al. [34] combined the opinion fusion model HK (Hegselmann-Krause, HK) and the epidemic transmission model SEIR (Susceptible-Exposed-Infected-Removed, SEIR) to explore the interaction behavior between users under interest and confidence thresholds. Zhao et al. [35] constructed an opinion-spreading model by combining the new coronavirus transmission model with the nature of information about public health emergencies, considering the important role of opinion leaders and the interest of individuals. This shows that this way of constructing information transmission models based on classical infectious disease models is effective.

Opinion control has been a key issue in the study of information spreading [36-39]. It is very dangerous to let completely unknown information spread freely in society and among the population. It is likely to mislead public opinion and affect economic development and the stability of social order [40, 41]. In recent years, many studies have attempted to minimize the negative impact of false or inaccurate information (rumors). Zhao et al. [42] reduced the impact of rumor generation by setting up hibernators (people-Hibernators) to connect the ignorant and the infected. Wang et al. [43] introduced discriminable and adversarial mechanisms to quantify people's level of cognitive ability and the competition between rumors and the truth to further curb the impact generated by rumors. Tripathy et al. [44] found that the delay of rumors is linearly related to the life cycle and concluded that the approach of embedding beacons in the network is more effective for rumor control.

### B. Motivations

* This work is supported by the National Natural Science Foundation of China under Grant 62266030 and Grant 61863025.(Corresponding author: Fuzhong Nian.)

1. The authors are with the College of Computer & Communication, Lanzhou University of Technology, Lanzhou 730050, China.

The above approaches are effective in reducing the scale of rumor spread and its impact on public opinion at the spreading level. However, a key issue is overlooked: the authenticity of the information is not clear until it has been confirmed. This means that it is inappropriate to control uncertain information by identifying it directly as a rumor. Therefore, this paper makes a clear distinction between two kinds of information based on the true spreading process. In this paper, information that has not been verified is defined as unconfirmed information. Information that has been proven to be true is defined as true information. Information that is explicitly proven to be false is defined as false information. Unconfirmed information can be logically converted into false information or true information.

Several studies have shown that individual behavior-based validation methods in the spreading process are often cumbersome and multidimensional [45, 46]. In order to facilitate the analysis of the intervention model, this paper simplifies the individual's own verification-seeking behavior and analyzes it by abstracting it into the impact that the announcement has on the individual. In reality, it is usually the government or authority that issues official notice to announce the authenticity of information. This type of behavior that can explicitly verify the authenticity of information is uniformly referred to as an announcement in this paper. This authentication mechanism is widely found in social networks, based on people's trust in authoritative organizations to carry out information release to complete the verification of a certain information. For example, the information released by official media such as People's Daily and CCTV to dispel rumors. In reality, in small groups, the information of group leaders can also play the role of information verification, such as department heads and heads of social groups. The control effect created by the announcement not only ensures the accuracy of the information, but also the correct direction of public opinion. Due to the potential differences in recommendation mode and identity verification among different social platforms, this paper focuses on analyzing the dissemination mode in the Sina Weibo platform.

Most of the existing studies on two-phase models in social networks focus only on the model's pattern and adaptability at the time of propagation, while the analysis of message authenticity and spread efficiency is lacking[47]. This paper divides the overall spreading process into two consecutive phases: the unconfirmed spreading phase and the confirmed spreading phase, using the intervention control of the announcement as the boundary. In the unconfirmed spreading phase, the unconfirmed information is disseminated like free spreading. When the announcement intervenes in the spreading process, the authenticity of the information is confirmed and the spreading enters the confirmed spreading phase. In this phase, the spreading speed is mainly controlled by the validation results.

Due to the highly heterogeneous nature of social networks, the node influence effect has also been the focus of many researchers' inquiry [48, 49]. The speed of spreading will be more affected if the information is trusted or questioned by those users who have more influence [50, 51]. In this paper, the influence of the node effect on the spreading process is simulated by building an incentive function.

*C. Task and Expected Results*

The main tasks of this paper are as follows: 1) The online social network model is established to simulate the real network at the network structure level; 2) The two-stage information spreading dynamics model based on announcement control was built. The control effect of announcements on the spreading process is analyzed, and the effectiveness is verified. 3) The spreading efficiency evaluation model is developed to analyze the effect of different announcement intervention times. The expected result of the work in this paper is to simulate the information spreading process for the experimental environment. And the experimental results will be compared with the real cases for trend comparison to verify the accuracy of the spread trend predicted by the model. The paper goes on to analyze, in depth, the role of announcement behavior in promoting true information and inhibiting false information. The visualization and comparison are also performed for the important parameters constructed in this paper to verify the validity of the theory and the reliability of the model.

*D. Technical Contributions*

The innovative technical contributions of this paper are as follows: 1) An innovative distinction is made between unverified information and rumors. And based on this, two stages of the spreading process are distinguished: the unverified spreading stage and the verified spreading stage. 2) This paper addresses the drawback of considering only the macroscopic role of spread in traditional studies by designing the role of announcements as an effect on individual nodes. 3) The intervention time of the announcement is analyzed in the simulation to extend the realistic meaning of the model.

*E. Organization*

The rest of the paper is organized as follows. The details of constructing the model and defining the important parameters are described in Section II. Among them, Part A and Part B constructed Online Social Network Model and Online Information Spreading Model for the influence of announcement control role, respectively. In Part C, Network State Transition Model is constructed to describe the state transition process within the network at the intervening moment. Part D constructs Two-stage Information Spreading Dynamics Model Based on Announcement Control on the basis of the previous model. Finally, the Spreading Efficiency Evaluation Model is constructed in Part E to realize the evaluation of the overall spreading efficiency. The results of the simulation experiments are presented and analyzed in part A of Section III. The results of the spreading efficiency evaluation model are analyzed in part B of Section III. The part C of Section III compares the experimental data with the real data and analyzes the strengths and weaknesses of the model in this paper. The part D of Section III presents experiments on the adaptive capacity of the model that are conducted to verify that the model can be effectively generalized to other spreading processes.

## II. MODEL BUILDING

### A. Online Social Network Model Based on Announcement Control

The conceptual diagram of the underlying network model for the experiments is shown in Fig. 1, and the building process is as follows.

1. Construct a BA (Barabasi-Albert) scale-free network with $N$ nodes as a global network $G(V,E)$, where $V=\{v_1,v_2,\cdots,v_N\}$ denotes the set of nodes of the network $G$ and $E=\{e_1,e_2,\cdots,e_N\}$ denotes the set of edges of network $G$.

2. Nodes are selected as initial known nodes in the global network with initial infection rate $i_0$ and redefined as sub-networks $G_I(V_I,E_I,C,R)$. $C=\{c_1,c_2,\ldots,c_m\}$ denotes the set of node confidence in the sub-network, where, $c_j \in [0,1]$. $R=\{r_1,r_2,\ldots,r_m\}$ denotes the set of node credibility in the sub-network. Its initial value $r_j=0$, while in the post-announcement network $r_j \in [-0.5, 0.5]$.

For the purpose of analysis, all known nodes in the network are defined as a sub-network $G_I$ in this paper. The set of nodes ($V_I$) of the sub-network contains all known nodes, and the set of edges ($E_I$) of the sub-network contains the edge relationships of these nodes in the global network. The difference with the global network is that each known node will get two new properties Confidence and Credibility, which form the sets $C$ and $R$, respectively.

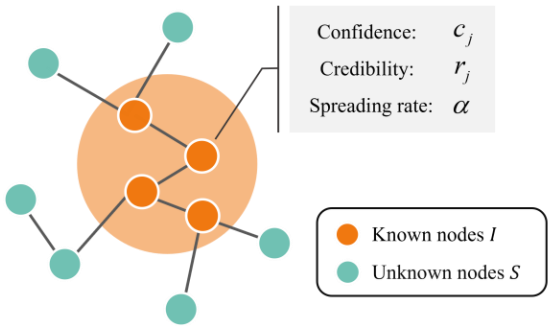

**Fig. 1.** Schematic of the online social network model based on announcement control. Each node in the sub-network consisting of I nodes has independent Confidence and Credibility properties and a spread rate $\alpha$ generated according to the model.

### B. Online Information Spreading Model Based on Announcement Control

The control role of the announcement is essentially the trust of individual nodes in their verification results. Therefore, the key issue of model building is the construction of node confidence and the correction of the state. The logic flowchart of the online information spreading model based on announcement control is shown in Fig. 2. In the unconfirmed spreading phase, the known nodes $I$ in the network can be distinguished according to the spreading attitude: known nodes that believe in the message content ($c_i > 0.5$), known nodes that do not believe in the message content ($c_i < 0.5$), and neutral known nodes ($c_i = 0.5$). They correspond to the green, red and yellow nodes in Fig. 2, respectively. Different spreading attitudes will lead to different spreading probabilities and spread to neighboring nodes.

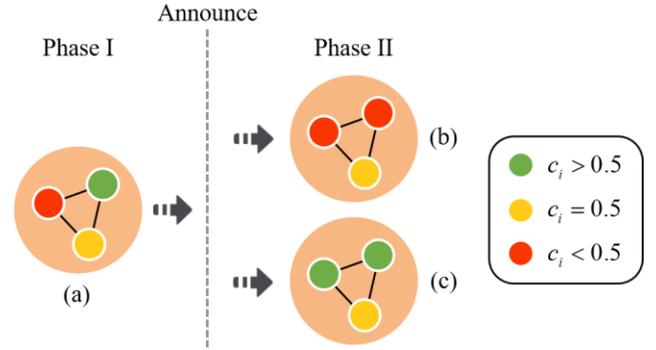

**Fig. 2.** Logic flowchart of online information spreading based on announcement control. There are three possible information states that exist in the spreading process. (a) Unconfirmed information state: the state of spreading when the information has not yet been validated. (b) False information state: the spreading state when the information has been verified as false. (c) True information state: the spreading state when the information has been verified as true.

When the announcement intervenes, the confirmed spreading phase is entered. At this time, node confidence and credibility are corrected under the effect of verification. In this phase, there are two possible forms of the composition of the spreading attitude in the network. When the prior unconfirmed information is proved to be false, only known nodes that do not believe the information content and neutral known nodes exist in the network, as shown in Fig. 2(b). When the prior unconfirmed information is proven to be true, only known nodes that believe the information content and neutral known nodes exist in the network, as shown in Fig. 2(c). After the correction, the information spreading is controlled (facilitated or inhibited) until the end of the spreading.

### C. Network State Transition Model

In the real world, when a message starts to spread early in the network, it is difficult for individual nodes to know its authenticity. The validation role of announcements in spreading enables individual nodes to quickly learn the authenticity of the information. In this paper, the network state transition model is developed to explain the impact of the validation role on nodes and the spreading process.

In this paper, the moment when the announcement intervenes in the spreading process is defined as the intervention point. The moment when the intervention occurs is denoted as $t_a$. For the sake of analysis, the ratio of the number of nodes in a state to the number of global nodes is referred to as the node density in this paper. At this point, the unconfirmed information in the first phase is confirmed. The moment when the difference in node density change is stable less than $5 \times 10^{-4}$ is noted as the end of spreading moment $t_f$. This means that in all cases constructed in this paper $t_f$ can be used as the time point at which the propagation trend stabilizes. To facilitate the analysis of the effect of intervention time, this paper defines the position of intervention $\tau$. Its value can be expressed as:

$$\tau = \frac{t_a}{t_f}. \quad (1)$$

It follows that the intervening position $\tau$ essentially defines the position of the intervening point in the global spreading, $\tau \in [0,1]$.

The intervention of validation will lead to two possible conclusions: the unconfirmed information is true or the unconfirmed information is false. In this paper, the information authenticity parameter is defined as $\beta$. In a complete spreading process, the information authenticity parameter determines the result of the unconfirmed information being verified. When $\beta = true$, the verification result is true information. When $\beta = false$, the verification result is false information.

In this paper, we assume that a message identified as a rumor will not be trusted by other individuals. The individual confidence and credibility are corrected under the global influence generated by the validation results. The specific procedure is shown in Algorithm 1: Node confidence correction. The algorithm calculates and reassigns the confidence and credibility of the nodes in response to different announcement results. Here *G* denotes the experimental network, *G.node[id][attributes]* denotes the attribute of the node object in the network, and *random* denotes the random number generation function.

**Algorithm 1: Node confidence correction.**

1: **if** $\beta$ = True **then**
2:    **for** all id in G **do**
3:       Dif = | G.node[id]['confidence'] - 0.5 |
4:       **if** G.node[id]['confidence'] < 0.5 **then**
5:          G.node[id]['credibility'] -= Dif
6:          G.node[id]['confidence'] = random (0.5~1)
7:       **else**
8:          G.node[id]['credibility'] += Dif
9:          G.node[id]['confidence'] += random (0~0.5)
10:         **if** G.node[id]['confidence'] > 1 **then**
11:            G.node[id]['confidence'] = 1
12: **if** $\beta$ = False **then**
13:    **for** all id in G **do**
14:       Dif = | G.node[id]['confidence'] - 0.5 |
15:       **if** G.node[id]['confidence'] > 0.5 **then**
16:          G.node[id]['credibility'] -= Dif
17:          G.node[id]['confidence'] = random (0~0.5)
18:       **else**
19:          G.node[id]['credibility'] += Dif
20:          G.node[id]['confidence'] -= random (0~0.5)
21:          **if** G.node[id]['confidence'] < 0 **then**
22:             G.node[id]['confidence'] = 0

*D. Two-stage Information Spreading Dynamics Model Based on Announcement Control*

In this paper, the SI model from the classical infectious disease model is selected as the base model. In the traditional SI model, an individual is infected and then remains infected forever [52]. Its main characteristics are: initially, some nodes are randomly defined as the infected state, and then the neighboring nodes of the initially infected nodes start to be infected. When a node is infected, it will remain in the infected state. Similarly, when events reached online, the earliest information is usually released by one or a few knowledgeable people, and then it develops into a large-scale spreading behavior. In the process of spreading, the information released by people has the characteristics of long-term retention and long-lasting influence. Even if individuals forget some of the information, it will not affect the content of their previous releases and the impact they caused. Therefore, based on the basic characteristics of the SI model, this paper establishes the two-stage information spreading dynamics model based on announcement control.

During the spreading process, the nodes that have been spread and acquired information are defined as known nodes $I$. The nodes that have not acquired information are defined as unknown nodes $S$. Thus, $S(t)$ and $I(t)$ denote the number of individuals with unknown information and the number of individuals with known information at moment $t$, respectively. Where $S(t) + I(t) \equiv N$. In the model of this paper, $\alpha$ denotes the probability that an unknown individual will come into contact with a known individual and be spread information per unit of time. The density of unknown and known nodes in the network at the moment $t$ can be expressed as:

$$\begin{cases} s(t) = \dfrac{S(t)}{N} \\ i(t) = \dfrac{I(t)}{N} \end{cases}, \quad (2)$$

where $s(t) + i(t) \equiv 1$, the basic spreading dynamics equation can be expressed as:

$$\begin{cases} \dfrac{ds}{dt} = -\alpha si \\ \dfrac{di}{dt} = \alpha si \end{cases}. \quad (3)$$

An individual's level of trust in an event is defined as the node confidence $c$. It is assumed that when a communicating individual acquires new information, he or she will spontaneously trust or question the information based on his or her perception. This in turn will lead to differences in individual spreading attitudes and affect the individual information spreading rate of the node.

In a realistic network system, nodes with larger degrees generally have more influence. For example, scholars prefer to cite papers by more established scholars to illustrate the results of their own work[53, 54]. In the dissemination process, this paper similarly analyzes the influence generated by nodes with larger degrees. In this paper, we establish a node influence incentive function based on node degree and node neighborhood to evaluate the influence of the node's degree on the spreading process. In the network model that has been developed above, the adjacency matrix can be expressed as $A = (a_{pq})_{N \times N}$. Then the degree of node $p$ in the network can be expressed as:

$$k_p = \sum_{q=1}^{N} a_{pq} = \sum_{q=1}^{N} a_{qp}. \quad (4)$$

The average degree of the global network can be expressed as:

$$\langle k \rangle = \frac{1}{N} \sum_{p,q=1}^{N} a_{pq}, \quad (5)$$

combining the node confidence and credibility, the information spreading rate of nodes can be defined as:

$$\alpha_p = \lambda_1 (c_p - \langle c \rangle) + \lambda_2 (r_p \cdot \frac{k_p - \langle k \rangle}{k_{\max}}). \quad (6)$$

Where $\lambda_1$ and $\lambda_2$ are the coefficients of confidence and credibility, respectively, $c_p$ and $r_p$ are the values of node confidence and credibility, respectively, $k_{\max}$ is the maximum degree in the global network, $\langle c \rangle$ denotes the neutral confidence. The first term of the equation analyzes the confidence of the nodes being spread, the second term analyzes the reputation and influence of the spread nodes. The node confidence reflects the trust level of the nodes being spread, and the higher the value, the higher the node spread rate, and vice versa, the lower the node spread rate. Credibility and influence will control the spread rate of a spreading node to its neighboring nodes, and obviously those nodes with higher influence and credibility values will have a stronger spread probability. The value of $\langle c \rangle$ can be expressed as:

$$\langle c \rangle = \begin{cases} \dfrac{1}{m} \sum_q c_q(t), & t \in [0, t_a), \; q \in V_I \\ 0.5, & t \in [t_a, \infty) \end{cases}, \quad (7)$$

where $t_a$ denotes the time when the intervention occurs and $m$ is the number of nodes in the sub-graph $G_I$. Thus, the logistic growth equation for the spreading process can be expressed as:

$$\frac{di}{dt} = \frac{1}{m} \sum_l [\lambda_1(c_l - \langle c \rangle) + \lambda_2 (r_p \cdot \frac{\sum_{q=p}^{N} a_{pq} - \frac{1}{N} \sum_{p,q=1}^{N} a_{pq}}{k_{\max}})] \cdot i(1-i), \quad (8)$$

where $l \in V_I$. And the solution can be obtained by mathematical derivation as:

$$i(t) = \frac{i_0 e^{\frac{1}{m} \sum_l [\lambda_1(c_l - \langle c \rangle) + \lambda_2 (r_p \cdot \frac{\sum_{q=p}^{N} a_{pq} - \frac{1}{N} \sum_{p,q=1}^{N} a_{pq}}{k_{\max}})] \cdot t}}{1 - i_0 + i_0 e^{\frac{1}{m} \sum_l [\lambda_1(c_l - \langle c \rangle) + \lambda_2 (r_p \cdot \frac{\sum_{q=p}^{N} a_{pq} - \frac{1}{N} \sum_{p,q=1}^{N} a_{pq}}{k_{\max}})] \cdot t}}, \quad (9)$$

where $i_0 = i(0)$. The specific procedure of the spreading process is shown in Algorithm 2.

---
**Algorithm 2: Information spreading process**

1. The number of $i_0 \cdot N$ nodes is selected as the initial known nodes, where $i_0$ denotes the initial infection density.

**I. The unconfirmed spreading phase:**

2. Generate node confidence $c \in [0,1]$, and assign it to the new known node $c_p$.

3. According to the attributes of individual nodes, the individual information spreading rate of nodes is calculated:

$$\alpha_p = \lambda_1(c_p - \langle c \rangle) + \lambda_2 (r_p \cdot \frac{\sum_{q=p}^{N} a_{pq} - \frac{1}{N} \sum_{p,q=1}^{N} a_{pq}}{k_{\max}}).$$

4. All known nodes are traversed, and its adjacent unknown node $S$ is changed to $I$ with probability $\alpha_p$.

5. Perform and repeat steps 2-4 until the announcement occurs, at which point spreading enters phase II. Perform global confidence correction (Algorithm 1) according to the authenticity of the information verified by the announcement.

**II. The confirmed spreading phase:**

6. According to the attributes of individual nodes, the individual information spreading rate of nodes is calculated $\alpha_p$.

7. All known nodes are traversed, and its adjacent unknown node $S$ is changed to $I$ with probability $\alpha_p$.

8. The node confidence generation rule at this stage is:

$$c \in \begin{cases} [0, 0.5], & \beta = false \\ [0.5, 1], & \beta = true \end{cases},$$

and assign it to the new known node $c_i$.

---

9. Perform and repeat steps 6-8 until spreading ends.

*E. Spreading Efficiency Evaluation Model*

For a message that is essentially true, the unconfirmed spread effectively facilitates the diffusion of this message. It improves the efficiency of information spreading, and at the same time, can promote the information to complete the global spreading earlier. However, for a message that is essentially false, the prior spread is not in the interest of society. The spreading process needs to be controlled as soon as possible. In addition to the prediction of the spread, the impact of the response cost of the announcement and the time-sensitive nature of the announcement needs to be considered.

The response cost of the announcement will result in that the cost of the announcement becomes higher as the response becomes faster. In real spread, the process of understanding the facts and generating decisions takes some time. Existing research suggests that the faster the response rate the more resources and costs are consumed[55, 56]. This is because rapid response requires a responsive information gathering department, efficient execution capabilities, and the ability to maintain focus. These qualities in reality often require excellent human resources and more exertion, which corresponds to a higher cost.

The time-sensitive nature of the information leads to a significant decrease in the strength of the announcement as time lags. Just as many studies have focused on the time-dependence of the value of information, i.e., the decrease in the value of information over time[57].

It is assumed that the earlier the intervention time, the higher the response cost and the stronger the announcement effectiveness. Conversely, the later the intervention time, the lower the response cost, and the weaker the announcement effectiveness. Then the response cost of the announcement and the time-sensitive nature of the announcement can be defined as: $\log_a \tau$ and $\log_b (1-\tau)$, respectively. In the model, the response cost function and the time-sensitive nature function of the announcement are controlled by parameters $a$ and $b$, respectively. In order to correctly predict the spreading effect due to different information authenticity, the evaluation model is divided into two parts. The evaluation function of the spread efficiency for true information can be expressed as Equation (10). Where $T_0$ denotes the total duration of the spreading process and $T_t$ denotes the duration required to reach the basic global information coverage ($i(t) \geq 0.95$) in the case of $\beta = true$.

$$E_t(\tau) = \frac{\varepsilon_1(\frac{T_0 - T_t(\tau)}{T_0})}{\log_a \tau + \log_b (1-\tau)}. \quad (10)$$

The evaluation function of the spread efficiency for false information can be expressed as:

$$E_f(\tau) = \frac{\varepsilon_2(\frac{I_0 - I_f(\tau)}{I_0})}{\log_a \tau + \log_b (1-\tau)}. \quad (11)$$

$I_0$ denotes the density of known nodes at the end of the free spreading state and $I_f$ denotes the density of known nodes at the end of the spreading in the case of $\beta = false$. For coefficients $\varepsilon_1$ and $\varepsilon_2$:

$$\varepsilon_2 = 1 - \varepsilon_1. \quad (12)$$

### III. SIMULATION AND ANALYSIS

This section shows the simulation results of online information spreading in online social networks to validate the theory in the previous section. The underlying experimental network in this paper is a network with N=2000 and <k>=5. To perform the modeling, a certain percentage of nodes need to be preset as the initial spreading nodes. These nodes are the first to publish information about the event and are the source of contagion at the beginning of the spreading process. The base conditions and the values of the key hyperparameters for the experiment are shown in TABLE 1.

TABLE 1
PARAMETER SETUP

| Symbol | Value | Definition |
|---|---|---|
| $N$ | 2000 | The total number of nodes |
| $k$ | 5 | Average degree |
| $i_0$ | 0.005 | The initial retweeting density |
| $\lambda_1$ | $3.875 \times 10^{-1}$ | Parameter of Eq.(5), the coefficients of confidence |
| $\lambda_2$ | $1.194 \times 10^{-1}$ | Parameter of Eq.(5), the coefficients of credibility |
| $\varepsilon_1$ | $8.12 \times 10^{-1}$ | Parameter of Eq.(10) |
| $\varepsilon_2$ | $1.88 \times 10^{-1}$ | Parameter of Eq.(10) |
| $a$ | $2.121 \times 10^{-1}$ | Control coefficient of time cost of announcement |
| $b$ | $3.089 \times 10^{-1}$ | Control coefficient of time-sensitive nature of announcement |

*A. Simulation Results*

Fig. 3 shows the change in the density of known nodes for three typical spreading scenarios when $\tau = 0.2$. The blue curve indicates the case when the message is not verified and spreads freely throughout. The red curve indicates the case when the unconfirmed information is verified as false news (rumor). The yellow curve indicates the case where the unconfirmed information is verified as true news. The spread is further stimulated when the intervention proves the message to be true. For example, the yellow curve in the figure reaches its peak about $t = 400$ earlier than the blue curve. However, when the intervention proves that the message is false, the spreading process is rapidly suppressed and the spreading speed is reduced to zero in a short time.

Fig. 4 shows the change curve of the density of known nodes for different intervention positions. The gray curve indicates the free spread case. The brown, yellow, green, red, and blue curves indicate the spreading when the values of the intervening positions are equal to 60%, 30%, 20%, 15%, and 10%, respectively. Fig. 4(a) shows the change in the density of known node when the information is verified as false. When the intervention position is less than 20%, the inhibition effect of the announcement is more obvious. The information inhibition rate basically reaches more than 20% and rises significantly as the intervention position becomes earlier. When the intervention position is greater than 30%, the inhibition effect decreases significantly. After the intervention position is greater than 60%, the change of communicator density is affected less, and the curve tends to be consistent with the free spreading situation. Fig. 4(b) shows the change of the density of known node when the information is verified as true. In this case, the original spreading process is stimulated when the intervention occurs. The information reaches global coverage at a faster rate.

Fig. 5 shows the visualization of the network structure in both cases. Where the green nodes indicate the unknown nodes $S$, the orange, red and yellow nodes all indicate known nodes $I$ in the spread process. The difference is that the orange nodes carry information that is unverified. The red and yellow nodes carry information that has been verified as false and true, respectively. The network structure at the beginning of the spread is shown in Fig. 5(a). A small number of informed nodes publish early information related to the event. The unconfirmed spreading phase begins. The information at this point is in the form of free spread because it has not been verified. The network structure at the end of spreading when the information is verified as false is shown in Fig. 5(b). As the information spreading rate is controlled in time after the announcement, the network structure at the end of spreading when the information is verified as true as shown in Fig. 5(c). Obviously, the validation effect of the announcement effectively motivates the information to spread. The unconfirmed information state in the first phase is converted into true information, and thus is propagated more rapidly. Eventually, the global information coverage is reached at a faster rate.

### B. Analysis of Information Spreading Efficiency

In order to effectively assess the effectiveness of the announcement intervention, this paper analyzes the efficiency of information dissemination at two main levels. At the information dissemination level, the effect of both true and false information verification results on the overall spreading process is analyzed. At the level of social reality, the evaluation of the response cost and information timeliness of the announcement is added. The simulation calculation process reduces the influence of random values on the overall process by means of repeated trials.

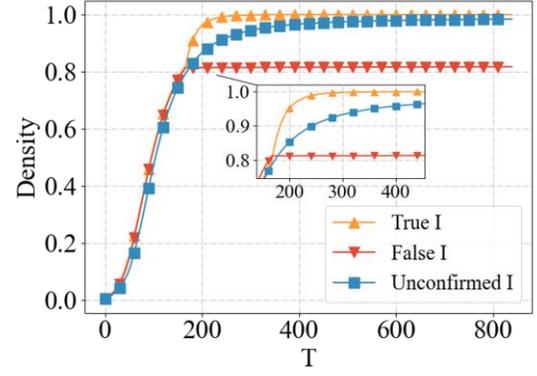

**Fig. 3.** Evolution of spreader node density for three typical spreading scenarios. The case where the message is always unverified, i.e., completely freely spread (blue). The case where an unconfirmed message is verified as a false message (red). The case where the unconfirmed message is verified as a true message (yellow).

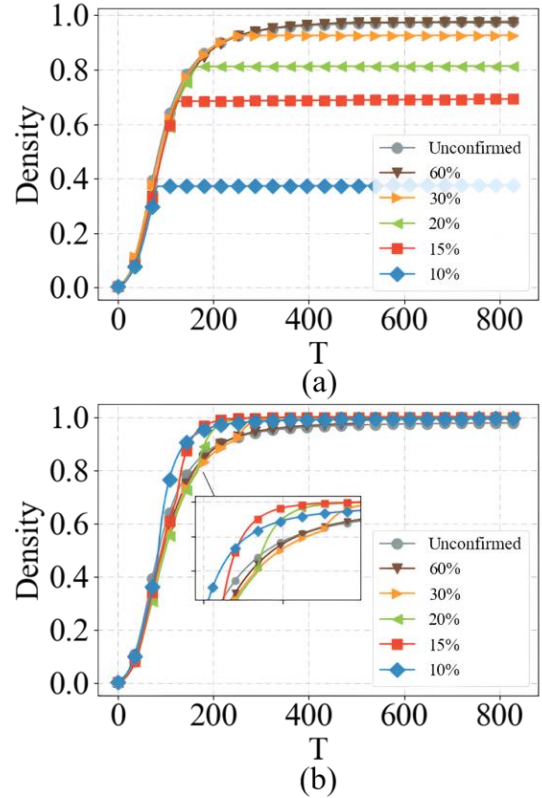

**Fig. 4.** Evolution of spreader density for different intervention positions. Different intervention positions: (a) Information is verified as false. (b) Information is verified as true.

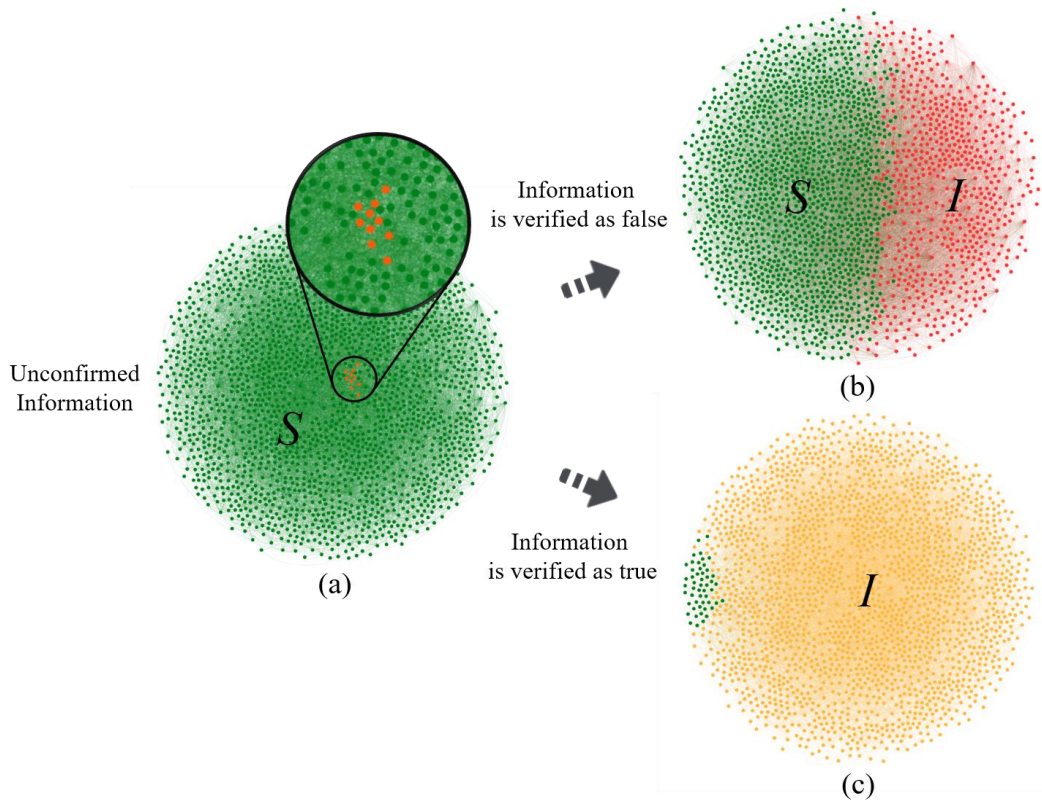

**Fig. 5.** Network evolution diagram. (a) Initial stage of network: a handful of informed nodes released early information related to the incident. (b) Final stage of network when information is verified as false: information dissemination is controlled in a timely manner. (c) Final stage of network when information is verified as true: rapidly achieve information coverage across the board.

The variation of efficiency scores is shown in Fig. 6. The horizontal coordinate is the position of intervention $\tau$ as defined previously. The vertical coordinate represents the assessment score results of the model. For ease of presentation, the graph indicates the intervention position in the form of a percentage. When the message is verified as true, the efficiency score changes as shown in the green curve. Due to the limiting effect of the response cost. The score is lower and in an increasing state when the intervention position is earlier and reaches a maximum at around $\tau = 20\%$. The low score at the beginning of the curve is due to the high response cost in a very short period of time. After that, the curve shows a surge because the response cost decreases gradually with time delay and the score increases rapidly. Due to the control effect of the time-sensitive nature. The efficiency score is in a state of rapid decline when spreading later. This is also consistent with the actual situation: after the message has been heavily discussed, the intervention effect of the announcement is greatly reduced. It is worth noting that the curves show significant non-monotonicity in the later stages, which this paper argues is due to the non-uniformity between single-step spreads. When the announcement occurs, if the new nodes include a larger proportion of large-degree nodes, then the promotion effect of the announcement will be more significant at this time than in other cases. When the information is verified to be false, the efficiency score changes as shown in the red curve. As can be seen from the figure, its maximum value occurs at around $\tau = 7\%$ compared to the green curve. This is due to the higher efficiency gain resulting from the earlier intervention position for the rumor spreading process. It also proves that when a rumor message is controlled early, it generates higher social benefits.

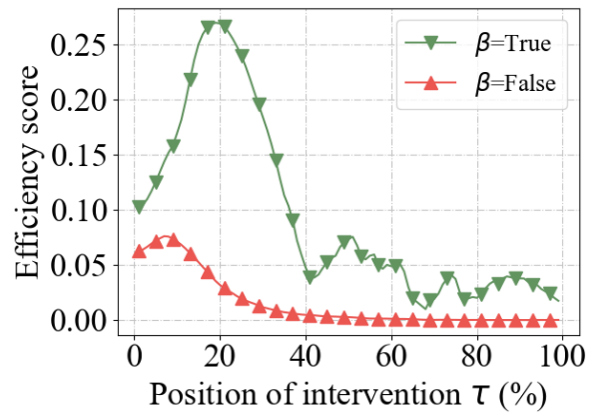

**Fig. 6.** Evolution of efficiency score. Efficiency score change curve (green) when information is verified as true. Efficiency score change curve (red) when information is verified as false.

*C. Contrast Analysis*

Information complex systems often lack guidance from first principles, so reasoning about the underlying principles from observed data is an important way to study these complex systems. In this section, the experimental results in the previous section are analyzed in comparison with the trends of the real data.

The real data in this paper are obtained from Sina Weibo. The data contain three sets of overall spreading processes of real events, which belong to three typical spreading cases: spreading verified as false, spreading verified as real and spreading without verification. One of the real data in Fig. 7(a-b) comes from the incident of a high school student falling from a building. The time span of the spread is from 11:20 on May 10 to 14:59 on May 11, 2021. More of the early information in this event eventually turns out to be false, which is a typical set of spreading processes that are verified as false. The real data in Fig. 7(c) are derived from an escape of animals that occurred in one location. The time span of the spread is from 13:44 on May 6 to 19:55 on May 25, 2021. This event was considered an ordinary accidental event in the early stage. However, the incident was revealed to be a major liability incident during the course of its spread. It is typical of a spreading process that is verified to be true. The real data in Fig. 7(d) are derived from the event of an actor's death in a car accident. The time span of the spread is from 17:30 on August 9 to 7:30 on August 10, 2021. There is no large-scale change in the overall content nature of the event. It is a typical unverified spread process.

In the real spread process, the overall scale of spread is influenced by many factors such as the nature of the event and the influence of the event. This leads to significant differences in the scale of the spread of different events. Therefore, the information spreading progress $R$ was used as the horizontal coordinate in the data visualization. In an attempt to reduce the interference due to scale differences, the spreading progress $R = T/T_0$, is expressed as a percentage. The announcement time parameter of the experiment is fixed at $\tau = 15\%$. A comparison of the density of known node change in the experimental data and the real data is shown in Fig. 7. Where the yellow curve indicates the spreaders density change in the real propagation case. The blue curve indicates the change of spreaders density in the experimental result data. For a more intuitive analysis, the bottom subplot shows the instantaneous growth rate changes of the corresponding curves.

The comparison of the density changes of the known nodes during the spreading process that was verified as false is shown in Fig. 7(a). In the simulation calculation, a part of unknown nodes is preserved in the network due to the effect of information suppression. The final spreader density is stabilized at around $i = 0.696$. And due to the limitation of obtaining data, it is difficult to obtain the information of unknown nodes in the real situation. Therefore, the experimental data here were normalized as shown in Fig. 7(b). This can effectively ignore the interference present in the experimental data and facilitate comparative analysis. In the former period of rapid spread, the experimental data and the real data trended in general agreement. When the message was verified to be false ($R = 15\%$), the message suppression effect was significant in both data sets. Message spreading was significantly controlled and terminated in a relatively short period of time.

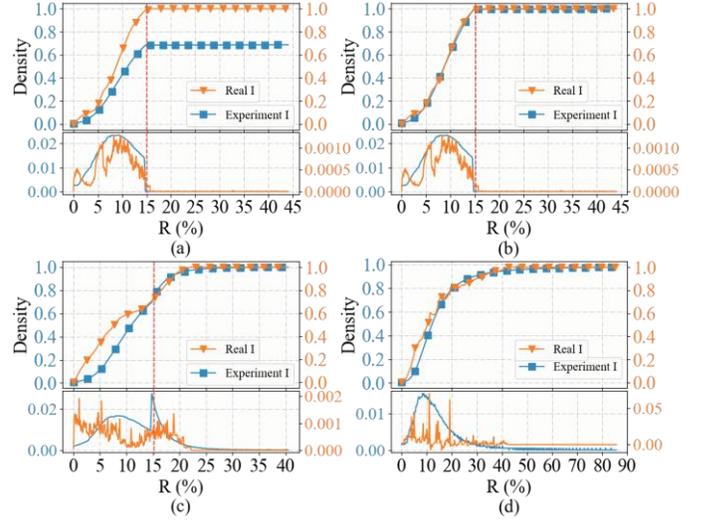

**Fig. 7.** Comparison of real cases and experiments. Three sets of typical reality cases used for comparison are from Sina Weibo: (a-b) Spread data verified as false. (c) Spread data validated as real. (d) Unconfirmed spreading data. The subplots at the bottom show the change in instantaneous growth rate of the corresponding curves.

The comparison of the density changes of the known nodes during the spreading process that was verified as real is shown in Fig. 7(c). Before the information was confirmed ($R \in [0,15]$), both the real situation and the experimental data showed free spread characteristics. The competition for information in the real spreading environment leads to a more pronounced decrease in the growth rate of spreaders in the real situation. After the information is confirmed ($R \in (15, \infty)$), there is an incentive increase in the density of spreaders of both real and experimental data. And the information coverage was largely completed after $R = 25\%$.

The comparison of the density changes of the known nodes during the spread without validation is shown in Fig. 7(d). In the absence of announcements for validation, the spread rate is mainly determined by the individual attributes of the nodes. The spreading process as a whole is characterized by free spreading. Compared to the spreading of the situation verified as true, it obviously takes longer spreading time ($R = 45\%$) to complete the information coverage.

*D. Model Adaptability*

In this section, the adaptability of the spread model developed in this paper is tested. Based on the previous analysis for the control efficiency, the announcement time of the experiment is set to $\tau = 0.15$ in order to ensure the experimental effect.

The variation of the density of known nodes, when spreading is performed in networks of different sizes is shown in Fig. 8(a). When the network size reaches 25 times the size of the experimental network (N=50,000), the spreading process still meets the expectation. The comparison of the spreading experiments using the WS (Watts-Strogatz) small world network as the base network is shown in Fig. 8(b). Since the WS network ignores more authority nodes (nodes with larger degree) in the construction process, compared with the BA network, the incentive effect generated by node influence is significantly limited in the WS network. This in turn leads to a significantly lower rate of rise for the yellow and red curves in the figure. And the validation effect of the announcement is still obvious.

The experiments with different network average degrees are shown in Fig. 8(c-d). Fig. 8(c) and Fig. 8(d) show the variation of known node density when the information authenticity parameters are $\beta = true$ and $\beta = false$ respectively. The network average degree limits the spread speed from the level of the number of neighbors of the nodes. The network average degree can limit the spread speed from the level of the number of neighbors of the nodes. Therefore, the increase in the average degree makes the spreading speed exhibit a small increase. But the basic trend of the spreading process remains consistent, since the node spread rate is mainly determined by the individual attributes of the nodes. This shows that the effect of announcement behavior on the spreading process is in line with expectations, including the spread excitation effect at $\beta = true$ and the spread suppression effect at $\beta = false$. The experiments with different initial infection rate conditions are shown in Fig. 8(e-f). Fig. 8(e) and Fig. 8(f) show the variation of known node density when the information authenticity parameters are $\beta = true$ and $\beta = false$ respectively. The initial infection rate can determine the number of known nodes in the initial spread. Therefore, when the initial infection rate increases, there is a small increase in the density of known nodes in the early stage. This also leads to a certain degree of weakening of the information suppression effect when $\beta = false$.

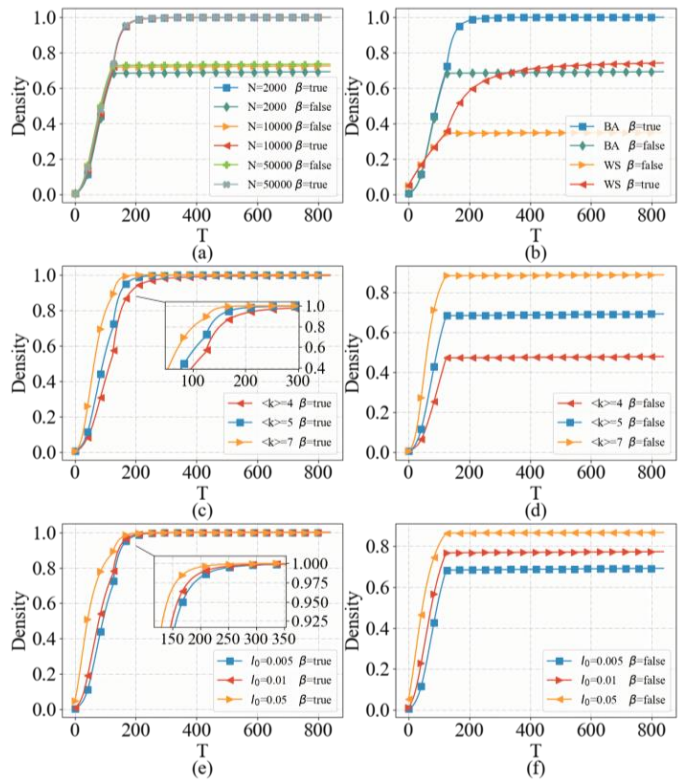

**Fig. 8.** Experiments on adaptability performance. Experiments on the generalization ability of the model are conducted in network environments of different sizes, types, averages, and initial spread densities.

The above experiments and analysis show that the model constructed in this paper can still show good results under different network environments and conditions. The feasibility of further extending the model to other spreading processes is ensured.

IV. CONCLUSION

This paper investigates the process of information spreading in modern online social networks and the influence of the control of announcements on the spreading process. The role of false information in the spreading process was explored by distinguishing it from true information. Two spreading stages were divided by the intervention of the announcement: the unconfirmed spreading phase and the confirmed spreading phase. The spread model was used as the basis for an in-depth analysis of the spread results due to the intervention time.

In this paper, an online social network model is first established based on the individual attributes of nodes. The exploration of node influence in traditional methods is abstracted as the role of node degree in the process of spreading. Meanwhile, the dynamic propagation rate of nodes is constructed by combining the individual attributes of nodes. In this way, the two-stage information spreading dynamics model based on announcement control is established. In addition, this paper establishes a spread efficiency evaluation model with response cost and announcement timing as important control factors. The impact of the timing of announcement intervention on the overall spreading process is simulated and analyzed.

By comparing the experimental data with the real cases, the rationality of the two spreading phases delineated in this paper is demonstrated. The three typical real spread cases of verification as false, verification as real and no verification were compared and analyzed separately. It is found that the simulated data basically conforms to the propagation trend of the real cases. Meanwhile, the adaptive capacity of the model is experimentally analyzed. The experiments show that the spread model established in this paper can simulate real spread cases. And it can be effectively generalized to other spreading processes. The research content and conclusions of this paper are expected to provide more scientific basis for the relevant departments or authorities to carry out the intervention behavior of information management control. At the same time, it provides effective ideas for the research related to information intervention control.

The analysis also reveals that there is still room for improvement of the work in this paper in the case of multiple messages competing for spread. In the future work, we will continue to promote the research in the direction of information spreading and explore the deeper spreading mechanism.

V. DATA AVAILABILITY STATEMENT

The data that support the findings of this study are available from WRD Big Data (https://research.wrd.cn/). Restrictions apply to the availability of these data, which were used under license for this study. Data are available from the authors upon reasonable request and with the permission of WRD Big Data. The experimental codes for this paper were done on "Pycharm" (https://www.jetbrains.com/pycharm). Using "Gephi" software (https://gephi.org/), the evolution of real network is visualized.

REFERENCES

[1] X. Wang, Y. Xing, Y. Wei, Q. Zheng, and G. Xing, "Public opinion information dissemination in mobile social networks–taking Sina Weibo as an example," *Information Discovery and Delivery,* 2020.
[2] D. Varshney and D. K. Vishwakarma, "A review on rumour prediction and veracity assessment in online social network," *Expert Systems with Applications,* vol. 168, p. 114208, 2021.
[3] M. E. Newman, "The structure and function of complex networks," *SIAM review,* vol. 45, no. 2, pp. 167-256, 2003.
[4] L. Zhu, G. Guan, and Z. Zhang, "Mathematical analysis of information propagation model in complex networks," *International Journal of Modern Physics B,* vol. 34, no. 26, p. 2050240, 2020.
[5] G. Guo, L. Ding, and Q.-L. Han, "A distributed event-triggered transmission strategy for sampled-data consensus of multi-agent systems," *Automatica,* vol. 50, no. 5, pp. 1489-1496, 2014.
[6] Y. Lou, D. Yang, L. Wang, C. Tang, and G. Chen, "Controllability robustness of Henneberg-growth complex networks," *IEEE Access,* vol. 10, pp. 5103-5114, 2022.
[7] C. Xia, Y. Luo, L. Wang, and H.-J. J. I. S. J. Li, "A fast community detection algorithm based on reconstructing signed networks," vol. 16, no. 1, pp. 614-625, 2021.
[8] Z.-K. Zhang, C. Liu, X.-X. Zhan, X. Lu, C.-X. Zhang, and Y.-C. Zhang, "Dynamics of information diffusion and its applications on complex networks," *Physics Reports,* vol. 651, pp. 1-34, 2016.
[9] Q. Wei, C.-j. Xie, H.-j. Liu, and Y.-h. Li, "Synchronization in node of complex networks consist of complex chaotic system," *AIP Advances,* vol. 4, no. 7, p. 077112, 2014.
[10] Y. Nie, S. Su, T. Lin, Y. Liu, and W. Wang, "Voluntary vaccination on hypergraph," *Communications in Nonlinear Science and Numerical Simulation,* vol. 127, p. 107594, 2023.
[11] Y. Nie, W. Li, L. Pan, T. Lin, and W. Wang, "Markovian approach to tackle competing pathogens in simplicial complex," *Applied Mathematics and Computation,* vol. 417, p. 126773, 2022.
[12] C. Dutta, G. Pandurangan, R. Rajaraman, and Z. Sun, "Information spreading in dynamic networks," *arXiv preprint arXiv:1112.0384,* 2011.
[13] X. Liu, D. He, and C. Liu, "Information diffusion nonlinear dynamics modeling and evolution analysis in online social network based on emergency events," *IEEE Transactions on Computational Social Systems,* vol. 6, no. 1, pp. 8-19, 2019.
[14] Y. Wang, K. You, M. Wang, Y. Huang, F. Chen, and C. Zhao, "Model of network community public opinion spread based on game theory," *Application Research of Computers,* vol. 30, no. 8, pp. 2480-2482, 2013.
[15] B. Wu, T. Yuan, Y. Qi, and M. Dong, "Public Opinion Dissemination with Incomplete Information on Social Network: A Study Based on the Infectious Diseases Model and Game Theory," *Complex System Modeling and Simulation,* vol. 1, no. 2, pp. 109-121, 2021.
[16] A. M. Kaplan and M. Haenlein, "Users of the world, unite! The challenges and opportunities of Social Media," *Business horizons,* vol. 53, no. 1, pp. 59-68, 2010.
[17] Z. Özbaş-Anbarlı, "Living in digital space: Everyday life on Twitter," *Communication & Society,* pp. 31-47, 2021.
[18] I. Leftheriotis and M. N. Giannakos, "Using social media for work: Losing your time or improving your work?," *Computers in Human Behavior,* vol. 31, pp. 134-142, 2014.
[19] A. Angali, M. Mojarad, and H. Arfaeinia, "ILSHR Rumor Spreading Model by Combining SIHR and ILSR Models in Complex Networks," *International Journal of Intelligent Systems & Applications,* vol. 13, no. 6, pp. 51-59, 2021.
[20] F. Nian and H. Diao, "A human flesh search model based on multiple effects," *IEEE Transactions on Network Science and Engineering,* vol. 7, no. 3, pp. 1394-1405, 2019.
[21] F. Nian, J. Ren, and X. Yu, "Online Spreading of Topic Tags and Social Behavior," *IEEE Transactions on Computational Social Systems,* 2023.
[22] D. Yang, T. W. Chow, L. Zhong, Z. Tian, Q. Zhang, and G. Chen, "True and fake information spreading over the Facebook," *Physica A: Statistical Mechanics its Applications,* vol. 505, pp. 984-994, 2018.
[23] A. Wonneberger and R. Vliegenthart, "Agenda-Setting Effects of Climate Change Litigation: Interrelations Across Issue Levels, Media, and Politics in the Case of Urgenda Against the Dutch Government," *Environmental Communication,* vol. 15, no. 5, pp. 699-714, 2021.
[24] J. Perego and S. Yuksel, "Media competition and social disagreement," *Econometrica,* vol. 90, no. 1, pp. 223-265, 2022.
[25] A. Castillo, J. Benitez, J. Llorens, and J. Braojos, "Impact of social media on the firm's knowledge exploration and knowledge exploitation: The role of business analytics talent," *Journal of the Association for Information Systems,* vol. 22, no. 5, pp. 1472-1508, 2021.
[26] Y. Nie, X. Zhong, T. Lin, and W. Wang, "Homophily in competing behavior spreading among the heterogeneous population with higher-order interactions," *Applied Mathematics and Computation,* vol. 432, p. 127380, 2022.
[27] M. Cinelli, G. De Francisci Morales, A. Galeazzi, W. Quattrociocchi, and M. Starnini, "The echo chamber effect on social media," *Proceedings of the National Academy of Sciences,* vol. 118, no. 9, p. e2023301118, 2021.
[28] J. Wang, L. Zhao, and R. Huang, "SIRaRu rumor spreading model in complex networks," *Physica A: Statistical Mechanics and its Applications,* vol. 398, pp. 43-55, 2014.
[29] A. Yang, X. Huang, X. Cai, X. Zhu, and L. Lu, "ILSR rumor spreading model with degree in complex network," *Physica A: Statistical Mechanics and Its Applications,* vol. 531, p. 121807, 2019.
[30] D. Yang, T. W. Chow, L. Zhong, and Q. Zhang, "The competitive information spreading over multiplex social networks," *Physica A: Statistical Mechanics its Applications,* vol. 503, pp. 981-990, 2018.
[31] Y. Zhu, Z. Zhang, C. Xia, and Z. J. A. Chen, "Equilibrium analysis and incentive-based control of the anticoordinating networked game dynamics," vol. 147, p. 110707, 2023.
[32] Z. Wang, Q. Guo, S. Sun, and C. Xia, "The impact of awareness diffusion on SIR-like epidemics in multiplex networks," *Applied Mathematics and Computation,* vol. 349, pp. 134-147, 2019.
[33] S. Kumar, M. Saini, M. Goel, and N. Aggarwal, "Modeling Information Diffusion In Online Social Networks Using SEI Epidemic Model," *Procedia Computer Science,* vol. 171, pp. 672-678, 2020.
[34] Q. Li *et al.*, "HK–SEIR model of public opinion evolution based on communication factors," *Engineering Applications of Artificial Intelligence,* vol. 100, p. 104192, 2021.
[35] J. Zhao, H. He, X. Zhao, and J. Lin, "Modeling and simulation of microblog-based public health emergency-associated public opinion communication," *Information Processing & Management,* vol. 59, no. 2, p. 102846, 2022.
[36] L. Yang, Z. Li, and A. Giua, "Containment of rumor spread in complex social networks," *Information Sciences,* vol. 506, pp. 113-130, 2020.
[37] L. Zhu, F. Yang, G. Guan, and Z. Zhang, "Modeling the dynamics of rumor diffusion over complex networks," *Information Sciences,* vol. 562, pp. 240-258, 2021.
[38] X. Yu, F. Nian, Y. Yao, and L. Luo, "Phase Transition in Group Emotion," *IEEE Transactions on Computational Social Systems,* vol. PP, no. 99, pp. 1-10, 2021.
[39] Z. Wang, C. Xia, Z. Chen, and G. Chen, "Epidemic propagation with positive and negative preventive information in multiplex networks," *IEEE transactions on cybernetics,* vol. 51, no. 3, pp. 1454-1462, 2020.
[40] S. Ai, S. Hong, X. Zheng, Y. Wang, and X. Liu, "CSRT rumor spreading model based on complex network," *International Journal of Intelligent Systems,* vol. 36, no. 5, pp. 1903-1913, 2021.
[41] C. Quan, L. Guowei, and L. Yiquan, "Research on the Modeling and Simulation of Network Public Opinion Evolution Considering User's Topic Interest Orientation," *Management Review,* vol. 32, no. 11, p. 128, 2020.

[42] L. Zhao, J. Wang, Y. Chen, Q. Wang, J. Cheng, and H. Cui, "SIHR rumor spreading model in social networks," *Physica A: Statistical Mechanics and its Applications,* vol. 391, no. 7, pp. 2444-2453, 2012.

[43] Y. Wang, F. Qing, J.-P. Chai, and Y.-P. Ni, "Spreading dynamics of a 2SIH2R, rumor spreading model in the homogeneous network," *Complexity,* vol. 2021, 2021.

[44] R. M. Tripathy, A. Bagchi, and S. Mehta, "A study of rumor control strategies on social networks," in *Proceedings of the 19th ACM international conference on Information and knowledge management*, 2010, pp. 1817-1820, 2010.

[45] R. Torres, N. Gerhart, and A. J. A. S. D. T. D. f. A. i. I. S. Negahban, "Epistemology in the era of fake news: An exploration of information verification behaviors among social networking site users," vol. 49, no. 3, pp. 78-97, 2018.

[46] R. Torres, N. Gerhart, and A. Negahban, "Combating fake news: An investigation of information verification behaviors on social networking sites," 2018.

[47] M. Azaouzi and L. B. Romdhane, "An efficient two-phase model for computing influential nodes in social networks using social actions," *Journal of Computer Science Technology,* vol. 33, pp. 286-304, 2018.

[48] M. Kitsak *et al.*, "Identification of influential spreaders in complex networks," *Nature physics,* vol. 6, no. 11, pp. 888-893, 2010.

[49] L. Lü, D. Chen, X.-L. Ren, Q.-M. Zhang, Y.-C. Zhang, and T. Zhou, "Vital nodes identification in complex networks," *Physics Reports,* vol. 650, pp. 1-63, 2016.

[50] J. Weng, E.-P. Lim, J. Jiang, and Q. He, "Twitterrank: finding topic-sensitive influential twitterers," in *Proceedings of the third ACM international conference on Web search and data mining*, 2010, pp. 261-270.

[51] S. Aral and D. Walker, "Identifying influential and susceptible members of social networks," *Science,* vol. 337, no. 6092, pp. 337-341, 2012.

[52] X.-F. Wang, X. Li, and G.-R. Chen, "Network science: an introduction," *Beijing: Higher Education Press,* vol. 4, pp. 95-142, 2012.

[53] D. W. Aksnes, "Characteristics of highly cited papers," *Research evaluation,* vol. 12, no. 3, pp. 159-170, 2003.

[54] C. Oppenheim and S. P. J. J. o. t. A. S. f. I. S. Renn, "Highly cited old papers and the reasons why they continue to be cited," vol. 29, no. 5, pp. 225-231, 1978.

[55] H.-W. Wang, S.-Y. Kuo, and L.-B. Chen, "Exploring the relationship between internal information security, response cost, and security intention in container shipping," *Applied Sciences,* vol. 11, no. 6, p. 2609, 2021.

[56] C. J. Pietras, A. E. Brandt, and G. D. Searcy, "Human responding on random‐interval schedules of response‐cost punishment: The role of reduced reinforcement density," *Journal of the Experimental Analysis of Behavior,* vol. 93, no. 1, pp. 5-26, 2010.

[57] A. Mohammadi, M. Saraee, and A. Mirzaei, "Time-sensitive influence maximization in social networks," *Journal of Information Science,* vol. 41, no. 6, pp. 765-778, 2015.

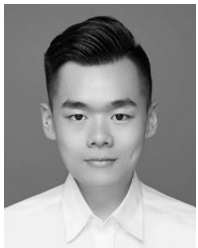

**Jinhu Ren** received the B.Eng. degree in of engineering from Shandong University of Science and Technology, Tai'an, China, in 2020, received the M.S. degree in computer technology from Lanzhou University of Technology, Lanzhou, China, in 2023, and currently pursuing the PhD in cyberspace security at the University of Science and Technology of China.

His main research interest includes the modeling and analysis of complex networks, with applications in information spread and human society.

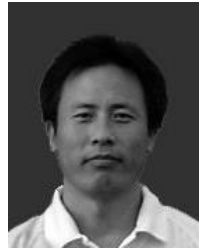

**Fuzhong Nian** received the B.S. degree in engineering from Northwest Normal University (department of Physics), Lanzhou, China, in 1998; the M.S. degree in engineering from Gansu University of Technology, Lanzhou, in 2004; and the Ph.D. degree in engineering from the Dalian University of Technology, Dalian, China, in 2011. He is interested in research at the intersection of mathematical modeling, network science, and control theory with application to biological, social, and chaotic network.

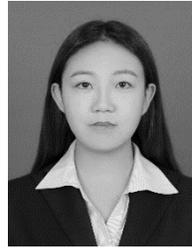

**Xiaochen Yang** received the B.Eng. degree in of engineering from LongDong University, Qingyang, China, in 2021, and is currently pursuing the M.S. degree in electronic information at Lanzhou University of Technology.

Her main research interest includes the modeling and analysis of complex networks, with applications to the direction of emotional contagion in society.